\documentclass[onecolumn,showpacs,preprintnumbers,a4paper,pre,notitlepage]{revtex4-1}
\usepackage{graphicx, subfigure}
\usepackage{amssymb, amsmath,amssymb,amsfonts}
\usepackage{dcolumn}
\usepackage{bm}
\usepackage{color}
\usepackage[utf8]{inputenc}

\begin{document}

\title{Tune the topology to create or destroy patterns}

\author{Malbor Asllani$^a$, Timoteo Carletti$^a$, Duccio Fanelli$^b$}
\affiliation{$^a$naXys, Namur Center for Complex Systems, University of Namur, rempart de la Vierge 8, B 5000 Namur, Belgium}

\affiliation{$^b$Dipartimento di Fisica e Astronomia, University of Florence, INFN and CSDC, Via Sansone 1, 50019 Sesto Fiorentino, Florence, Italy}

\begin{abstract}
We consider the dynamics of a reaction-diffusion system on a multigraph. The species share the same set of nodes but can access different links to explore the embedding spatial support. By acting on the topology of the networks we can control the ability of the system to self-organise in macroscopic patterns, emerging as a symmetry breaking instability of an homogeneous fixed point. Two different cases study are considered: on the one side, we produce a global modification of the networks, starting from the limiting setting where species are hosted on the same graph. On the other, we consider the effect of inserting just one additional single link to differentiate the two graphs. In both cases, patterns can be generated or destroyed, as follows the imposed,  small, topological perturbation.  Approximate analytical formulae allows to grasp the essence of the phenomenon and can potentially inspire innovative control strategies to shape the macroscopic dynamics on multigraph networks. 
\end{abstract}

\pacs{89.75.Hc, 89.75.Kd, 89.75.Fb}

\maketitle

\section{Introduction}
\label{sec1}

Macroscopic collective behaviours do emerge spontaneously in systems constituted by many-body interacting entities. This is a widespread observation in nature with many interdisciplinary applications ranging from biology to physics. Elucidating the key processes yielding to macroscopically ordered patterns is hence a fascinating field of investigations, at the forefront of many exciting developments. The mathematics that underlies patterns formation focuses on the dynamical interplay between reaction and diffusion processes.  Irrespectively of the specific domain of applications, elementary constituents can be ideally grouped in distinct species, family of homologous interacting units. Usually, reaction-diffusion models are defined on a regular lattice, either continuous or discrete. In many cases of interest, it is however more natural to place the system on a network, bearing a complex structure. Patterns for multi-species reaction-diffusion systems defined on complex networks materialise in a spontaneous differentiation between activator(inhibitor)-rich and activator(inhibitor)-poor nodes~\cite{nakao}. Directed coupling can further seed topologically driven patterns, for a choice of the reaction parameters for which the trivial homogenous solution proves stable~\cite{directed}.  Single individual effects are also crucial and significantly modify the idealised mean-field predictions: the stochastic component of the microscopic dynamics resulting from the inherent discreteness of the system, can in fact induce regular macroscopic patterns, both in time and space~\cite{stoch,asllani}.  

Self-organisation may however proceed across interlinked networks, by exploiting the multifaceted nature of resources and organisational skills. To account for the hierarchical organisation in multiple nested layers, networks of networks can be also considered. These concepts are particularly relevant to transportation systems~\cite{OlfatiSaberMurray2004,JohnDusiClaffy2010}, the learning process in the brain~\cite{SpornsTononiKootter2005} and to understanding the emergent dynamics in ecology and social communities in general~\cite{RietkerkvandeKoppel2008}.
The process of pattern formation for a generalised reaction-diffusion scheme hosted on multiplex has been studied in~\cite{multiplex}. Depending on the cooperative interference between adjacent layers, stratified patterns can emerge also when the deterministic instability on each individual layer is impeded. Conversely, patterns may dissolve by properly tuning the degree of interlayer overlap. Diffusion-induced instability have been also studied for reaction systems defined on multi-graphs, graphs decorated with multiple links between pairs of nodes~\cite{bollobas}, and shown to bear peculiar traits~\cite{kouvaris}. In all considered cases, the topology of the spatial support impacts on the ability of the systems to yield macroscopically organised patterns.

The aim of this paper is to expand on these ideas, bridging the frameworks discussed in~\cite{multiplex, kouvaris} and providing further evidence on the key role played by topology in shaping the system response to an external perturbation. More specifically, we will focus on the simplified setting where just two alternative sets of links are assumed to connect the available nodes. Each species can diffuse from one node to the other, engaging only one of the two distinct transportation layers. These latter are characterised in terms of their associated (weighted) adjacency matrices, which depend parametrically on a scalar quantity $\epsilon$. When $\epsilon=0$ the two graphs are identical. At variance, for $\epsilon=1$, the graphs are independent complex networks generated via an assigned recipe. By continuously increasing $\epsilon$, within the allowed interval of definition, one can access intermediate configurations. As we shall make clear in the following, the amount of disorder imposed at the scale of individual layers (and, consequently, their respective degree of diversity) can be modulated to effectively control the large scale dynamics of the scrutinised system. Disorder and diversity, as encoded in the scalar factor $\epsilon$, can make the patterns to emerge or, alternatively, fade away. Analytical estimates for the critical $\epsilon$ are obtained by perturbatively characterising the spectrum of the multi-dimensional matrix that governs the linear dynamics of the system, close to the homogeneous solution. Patterns obtained for increasing  $\epsilon$ appear to progressively localise~\cite{vanag,koga} on a subset of nodes, a phenomenon that eventually reflects the topological characteristics of the dominant eigenmode. In the second part of the paper, we consider an alternative formulation of the problem. The graphs that define the layers of the multigraph differ now by a single undirected edge, with given weight $w$. By tuning $w$, one can control the onset of the instability, as we shall prove analytically. 
Different topological controllers that act on the structural parameters of hosting network (nodes and edges), and their associated characteristics (weights), can be hence devised which interfere with the inherent ability of the system to self-organise in macroscopically ordered patterns. It is the heterogenous nature of the spatial support, and the non trivial coupling between adjacent layers, which instigate (or deter) the instability, an observation that can be in turn exploited to alter the fate of the system, without touching at its internal reactive dynamics, and thus providing a control strategy~\cite{control1, control2} for the onset or disruption of macroscopic patterns on complex networks.
 
The paper is organised as follows: in the next section the mathematical formalism is presented. In particular, we carry out a perturbative study of the spectrum of the multi-dimensional Jacobian matrix which governs the evolution of the perturbation, under the linear approximation. The issue of pattern localisation is also discussed. In Section 3, we make the two layers of the multi-graph distinct through insertion of a single additional 
link and provide analytical formulae to quantify its impact on the overall dispersion relation. The weight of the introduced link serves as a small parameter in carrying out a perturbative expansion. Finally, we sum up and conclude.

\section{Reaction-diffusion equations for multigraph networks}
\label{sec2}

\begin{figure}[ht!]
\begin{center}
\includegraphics[width=0.7\textwidth]{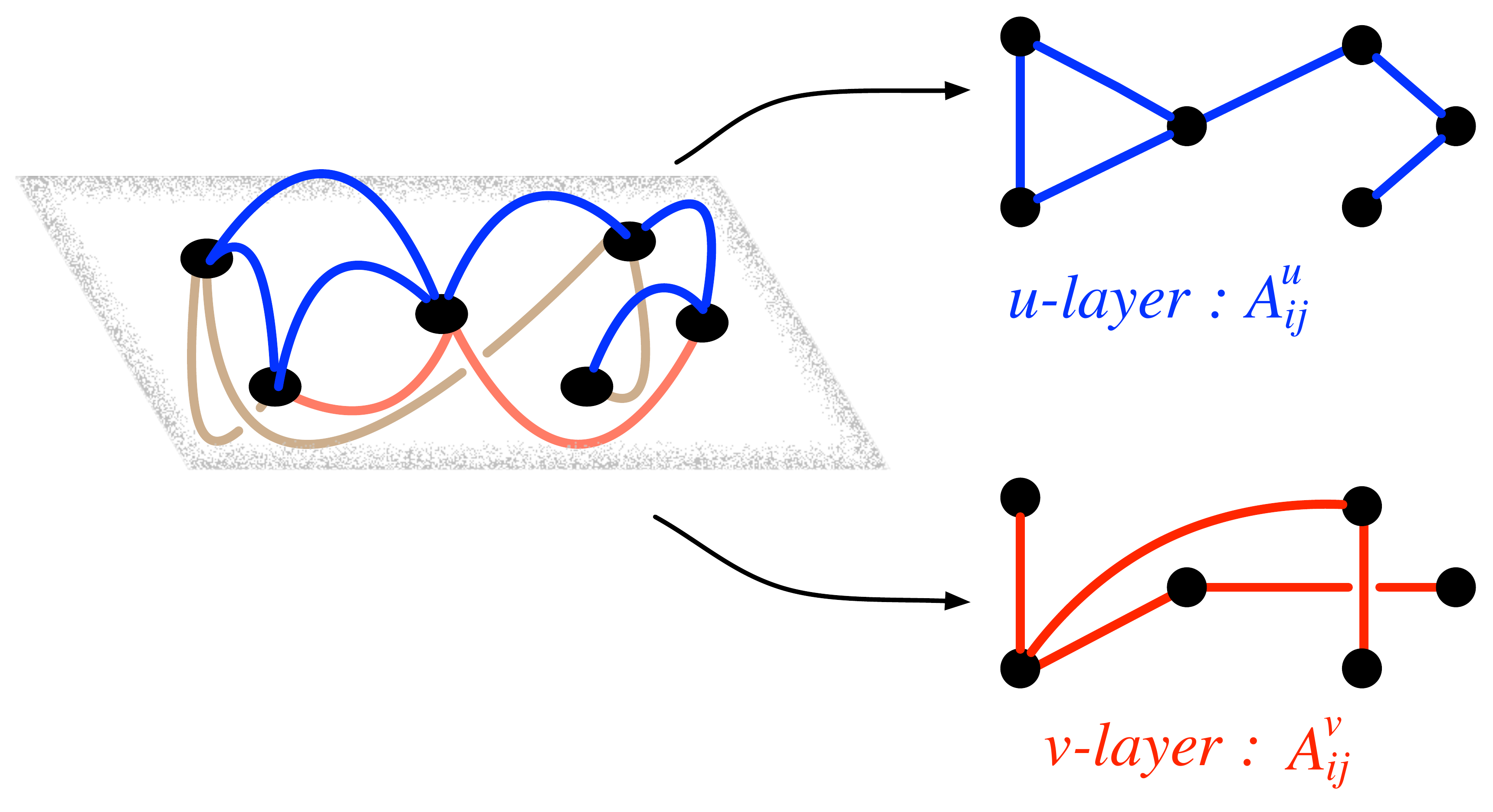}
\end{center}
\caption{A cartoon of a multigraph on the top of which the reaction-diffusion system evolves. The black circles denotes the nodes common to both species, blue links denotes the channels through which species $u$ can move, while red links are the ones for the $v$ species.}
\label{fig:multigraph}
\end{figure}

Imagine two species to interact and diffuse on a multigraph~\cite{bollobas}. For a sake of simplicity, we assume that each pair of nodes can be connected at most by a double link. Label with $u_i$ and $v_i$ the densities of the species on node $i$, and denote with $\Omega$ the total number of  nodes. The species undergo local reactions, via a standard activator-inhibitor scheme, and can diffuse among the nodes through the connecting links that are made available in the layer they belong to (see Fig.~\ref{fig:multigraph} for a schematic representation). 

Mathematically, we can cast the model in the following form:   
\begin{eqnarray}
\dot{u}_i&=& f(u_i,v_i) + D_u\sum_{j=1}^{\Omega}{L}_{ij}^u u_j \nonumber\\
\dot{v}_i&=& g(u_i,v_i) + D_v\sum_{j=1}^{\Omega}L_{ij}^v v_j
\label{eq:reac_diff}
\end{eqnarray}
where $f$ and $g$ refer to the nonlinear reactions, $\textbf{L}^u$, (resp. $\textbf{L}^v$) indicates the Laplacian matrix for the undirected network which defines the heterogeneous spatial support accessible to species 
$u$ (resp. $v$). More specifically, label with $A_{ij}^x$ the (symmetric and weighted) adjacency matrix of the network explored by species $x=u,v$. Then the Laplacian operator reads $L_{ij}^x=A_{ij}^x-k_i^x\delta_{ij}$, where $k_i^x=\sum_j A_{ij}^x$ is the degree of node $i$, referred to the layer  $x$ \cite{nakao}. In the following we shall illustrate our results with reference to the celebrated Brusselator model. This amounts to setting  $f(u,v)=1-(b+1)u+cu^2v$ and $g(u,v)=bu-cu^2v$ where $b$, $c$ are parameters of the model. The methodologies developed are however general and transcend the specific application here considered. Summing up,  Eqs.~\eqref{eq:reac_diff} define the general reaction-diffusion system on a multigraph network that we shall inspect in the rest of the paper. This formulation of the process was first proposed in \cite{kouvaris}. In the following, we will discuss the conditions that underly the instability, by adapting to this reference framework  the analytical techniques developed in \cite{multiplex}. 

\subsection{The linear stability analysis: a perturbative approach}

To determine the possible onset of the instability, following the scheme pioneered by A. Turing \cite{turing}, one has to preliminary require the existence of a stable homogeneous equilibrium, namely $u_i=u^*$ and $v_i=v^*$ for all $i=1,\dots,\Omega$. Non homogenous perturbation can turn unstable, as follows a symmetry breaking instability that reflects the non trivial interplay between reaction and diffusion terms. To shed light onto this issue, it is customary to perform a linear stability analysis of the non linear model~\eqref{eq:reac_diff}. By setting 
$u_i=u^*+\delta u_i$, $v_i=v^*+\delta v_i$ and linearising for small perturbation eventually yields: 
\begin{equation}
\left( \begin{array}{ccc}
\dot{\delta\boldsymbol{u}}\\\dot{\delta\boldsymbol{v}}
 \end{array} \right)=\left( \begin{array}{ccc}
f_u \mathbf{I}_\Omega + D_u\boldsymbol{L}^u & f_v \mathbf{I}_\Omega\\
g_u \mathbf{I}_\Omega & g_v \mathbf{I}_\Omega + D_v\boldsymbol{L}^v
 \end{array} \right)\cdot\left( \begin{array}{ccc}
\delta\boldsymbol{u}\\\delta\boldsymbol{v}
 \end{array} \right)=:\boldsymbol{\mathcal{\tilde{J}}}\left( \begin{array}{ccc}
\delta\boldsymbol{u}\\\delta\boldsymbol{v}
 \end{array} \right)
 \label{eq:lin_prob}
\end{equation}
where $\textbf{I}_\Omega$ is the $\Omega \times \Omega$ dimensional identity matrix. The matrix $\boldsymbol{\mathcal{\tilde{J}}}$  is the generalised Jacobian and  it is formally defined via the latter equality: $f_u$, $f_v$, $g_u$ and $g_v$ stand for the partial derivatives of the reaction terms, evaluated at the homogeneous equilibrium $(u^*, v^*)$.

The stability of the null solution of~\eqref{eq:lin_prob} can be assessed by solving the eigenvalue problem of the $2\Omega \times 2\Omega$ matrix $\boldsymbol{\mathcal{\tilde{J}}}$. In general this task cannot be achieved analytically and one has to resort to numerical methods. Our aim is to show that suitable perturbative techniques can be implemented to progress in the analytical characterisation of the conditions that underly the instability. More specifically, we assume that  the adjacency matrices that characterises the independent layers of the multigraphs can be respectively written as $A^u(\epsilon)=A^0+\epsilon\left(A^1-A^0\right)$  and $A^v(\epsilon)=A^0+\epsilon\left(A^2-A^0\right)$, where $\epsilon \in [0,1]$. 
$A_0$ specifies the topological characteristics of a network that the two species happen to share when $\epsilon = 0$. Conversely, for $\epsilon \ne 0$,  species relocate in space following distinct routes, the differences being more pronounced as  $\epsilon$ approaches unit. $A^1$ and $A^2$ identify the graphs made available to species $u$ and $v$,
when  $\epsilon = 1$. Starting from this setting, we will carry out a perturbative study of the spectral properties of matrix  $\boldsymbol{\mathcal{\tilde{J}}}$ , using $\epsilon$ as a small parameter in the expansion. By enhancing the degree of diversity among layers, via $\epsilon$, one can either instigate or silence the instability, thus controlling the route towards the subsequent pattern formation. 

Let us denote by $\textbf{L}^0$ the Laplacian matrix of the \lq\lq unperturbed\rq\rq network $A^0$. Then $\boldsymbol{L}^u=\boldsymbol{L}^0+\epsilon(\boldsymbol{L}^1-\boldsymbol{L}^0)$ and $\boldsymbol{L}^v=\boldsymbol{L}^0+\epsilon(\boldsymbol{L}^2-\boldsymbol{L}^0)$. Hence we can rewrite the matrix $\boldsymbol{\mathcal{\tilde{J}}}$ as follows
\begin{equation}
\boldsymbol{\mathcal{\tilde{J}}}=\left( \begin{array}{ccc}
f_u \mathbf{I}_\Omega + D_u\boldsymbol{L}^0 & f_v \mathbf{I}_\Omega\\
g_u \mathbf{I}_\Omega & g_v \mathbf{I}_\Omega + D_v\boldsymbol{L}^0
 \end{array} \right)+\epsilon\left( \begin{array}{ccc}
D_u\left(\boldsymbol{L}^1-\boldsymbol{L}^0\right) & \mathbf{0}_\Omega\\
\mathbf{0}_\Omega & D_v\left(\boldsymbol{L}^2-\boldsymbol{L}^0\right)
 \end{array} \right)=\boldsymbol{\mathcal{\tilde{J}}}_0+\epsilon\boldsymbol{\mathcal{D}}_0\, ,
 \label{eq:pert_J}
\end{equation}
where $\boldsymbol{\mathcal{D}}_0$ is defined by the last equality. Consider $\epsilon>0$ to be small. One can therefore approximate the spectrum of 
$\boldsymbol{\mathcal{\tilde{J}}}$ as a  local perturbation of the spectrum of $\boldsymbol{\mathcal{\tilde{J}}}_0$, which is assumed known a priori. The method of analysis that 
we shall adopt follows the ideas developed in~\cite{multiplex} and detailed in the annexed Appendix~\ref{appendice1}.

Let us denote by $\lambda^{(0)}_{\alpha}$  the eigenvalues of $\boldsymbol{\mathcal{\tilde{J}}}_0$ for $\alpha=1,\dots,2\Omega$; $\psi^{(0)}_{\alpha}$ and $\phi^{(0)}_{\alpha}$ stand for the corresponding right and left eigenvectors. Let us also assume, for simplicity, that the eigenvalues of $\boldsymbol{\mathcal{\tilde{J}}}_0$  are distinct: we have hence a set of linearly independent eigenvectors. This latter assumption can be 
relaxed, yielding more cumbersome computations that, however, do not add any further insight to the problem at hand~\cite{multiplex}.

Let $\tilde{\lambda}_{max}$ be the eigenvalue of $\boldsymbol{\mathcal{\tilde{J}}}$ with the largest real part. Assuming that the relative order of the eigenvalues is not affected by the imposed perturbation (otherwise one can easily compensate for such an effect), it is possible to trace back  $\tilde{\lambda}_{max}$ to the corresponding quantity  $\lambda^{(0)}_{max}$,  the eigenvalue of $\boldsymbol{\mathcal{\tilde{J}}}_0$ with the largest real part. By denoting with  $\psi^{(0)}_{max}$ and $\phi^{(0)}_{max}$ the right and left eigenvectors of 
 $\boldsymbol{\mathcal{\tilde{J}}}_0$  relative to $\lambda^{(0)}_{max}$, one can write:

\begin{equation}
\tilde{\lambda}_{max}=\lambda^{(0)}_{max}+\epsilon\lambda_{max}^1+\epsilon^2\lambda_{max}^2+\mathcal{O}(\epsilon^3)=\lambda^{(0)}_{max}+\epsilon\frac{{(\phi^{(0)}_{max})^T}\boldsymbol{\mathcal{D}}_0\psi^{(0)}_{max}}{(\phi^{(0)}_{max})^T\cdot \psi^{(0)}_{max}}-\epsilon^2  \frac{(\phi^{(0)}_{max})^T \boldsymbol{\mathcal{D}}_0\psi^{(1)}_{max}}{(\phi^{(0)}_{max})^T\cdot \psi^{(0)}_{max}}+\mathcal{O}(\epsilon^3)\, ,
\label{eq:pert_lambda}
\end{equation}
where the definition of $\psi^{(1)}_{max}$ is given in the Appendix~\ref{appendice1}, together with a detailed derivation of the above formula.
Observe that Eq.~\eqref{eq:pert_lambda} contains first and second order corrections in $\epsilon$: high order corrections can be also computed, following the procedure described in Appendix \ref{appendice1} (see also Fig.~\ref{fig:Fig1}). 

The approximated expression (\ref{eq:pert_lambda}) opens up the perspective to assess the stability of the system versus $\epsilon$, for any choices of the arbitrary adjacency matrices $A^0$,  $A^1$ and $A^2$. Assume for instance that patterns cannot develop for $\epsilon=0$, namely when species $u$ and $v$ diffuse on the same network, as 
specified by the adjacency matrix  $A^0$. This in turn implies $\Re\lambda^{(0)}_{max}<0$, for $\epsilon=0$ ($\Re (\cdot)$ selects the real part of $(\cdot)$). This condition is met in Fig.~\ref{fig:Fig1}(a), for the Brusselator model. Here, $A^0$ identifies a regular one dimensional lattice, with long-range links. In this example, we solely modify the network of the inhibitor species. This latter is in fact made to evolve on a graph which results from the linear combination of the lattice $A^0$ and a Watts-Strogatz (WS) network $A^2$.  As it is evident by direct inspection of Fig.~\ref{fig:Fig1} (a),  a modest diversification of the networks hosted on each layer  suffices to drive the instability and opens the route to patterns development. The opposite scenario also holds true, as illustrated in 
Fig.~\ref{fig:Fig1}(b). Now the patterns are allowed when the species live on a shared regular lattice. If the activator network is modified via $A^1$, a random graph generated with the Watts-Strogatz recipe, the system recovers stability for a suitable choice of the coupling parameter $\epsilon$.  The accuracy of the analytical approximation sensibly depends on the inspected setting. In all cases, however, the theory enables us to foresee if the networks contamination will eventually drive the system unstable or if, conversely, it will restore the stability. Approximate values for the critical threshold $\epsilon_c$ can be also obtained, as a straightforward consequence.

As an important remark, we notice that in deriving Eq.~\eqref{eq:pert_lambda} use has been made of the eigenvectors of the matrix $\boldsymbol{\mathcal{\tilde{J}}}_0$. Alternatively, we could have employed the eigenbasis of $\boldsymbol{L}_0$ yielding a formula similar to that  proposed by~\cite{nakao}. In both cases, however, one has to study the spectral properties of a  $2 \Omega \times 2 \Omega$ matrix, a task that requires resorting to numerical methods.

\begin{figure}[ht!]
\begin{center}
\subfigure[]{
\includegraphics[width=0.4\textwidth]{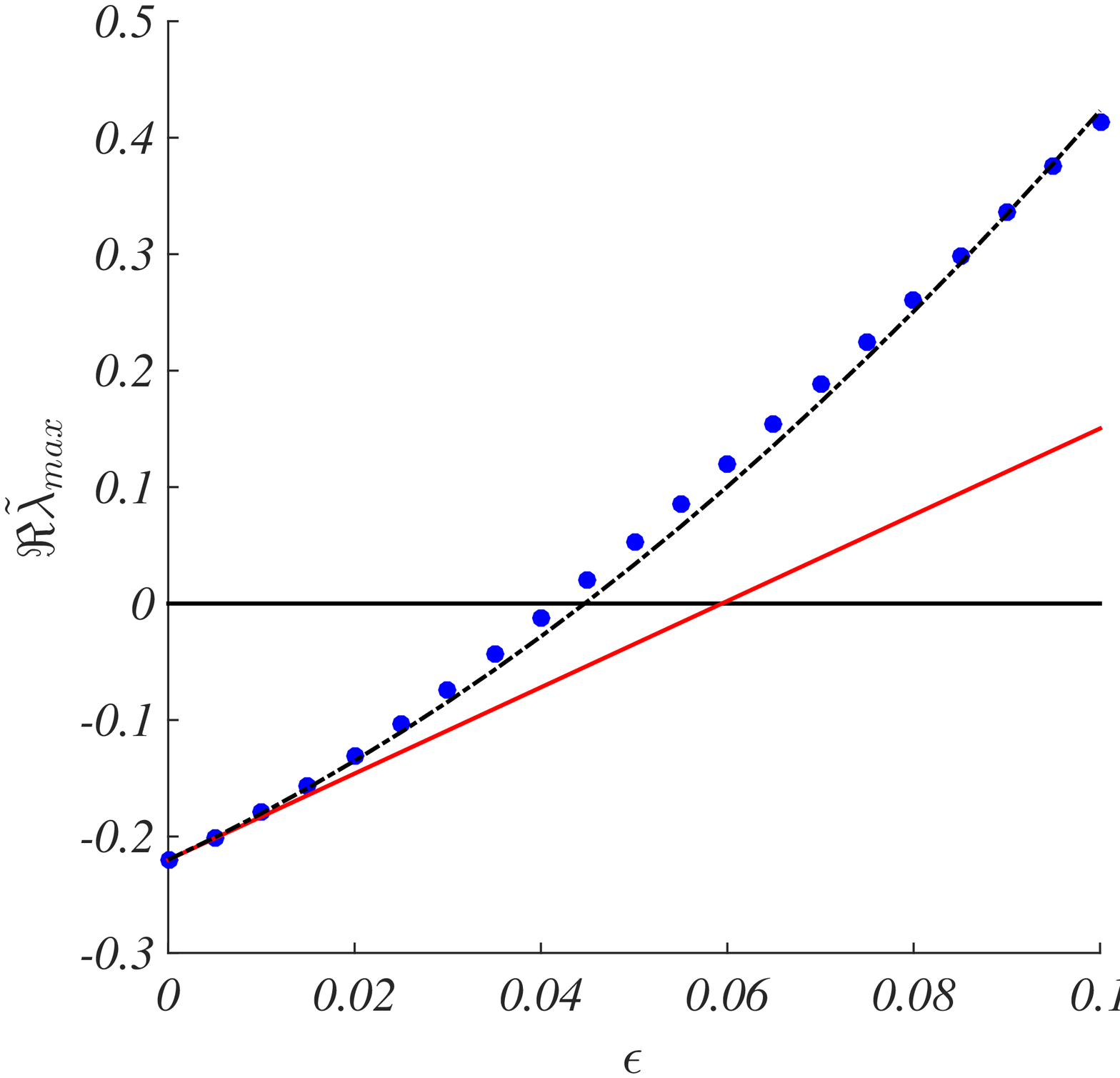}
}\hspace*{-0.5cm}
\subfigure[]{
\includegraphics[width=0.4\textwidth]{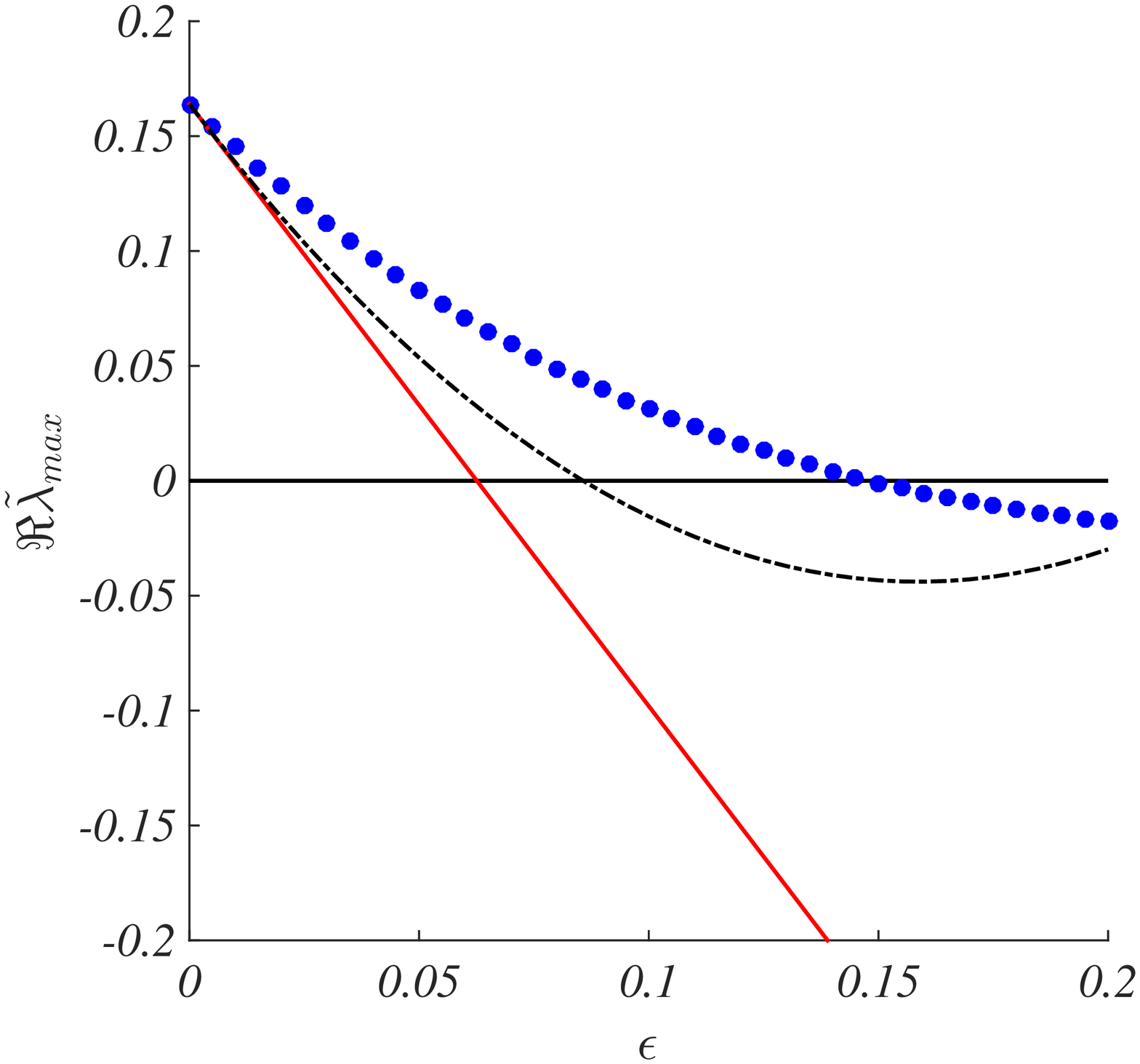}
}
\end{center}
\caption{Dispersion relation for the multigraph. Panel a). Patterns do not exist when both species use the same network $A^0$ (a ring made by $N=100$ nodes: each node is linked to its $3$ closest neighbours on the left and on the right) while they emerge once the inhibitor is allowed to use different paths $A^v(\epsilon)=A^0+\epsilon(A^2-A^0)$, $\epsilon>0$, where $A^2$ is a Watts-Strogatz~\cite{WS1998} networks obtained using $N=100$ nodes, $\langle k\rangle=6$ and a probability to rewire of $p=0.8$. The activator species is always made to evolve on $A^0$. The reaction terms follow from the Brusselator model with parameters $b=8$, $c=26$, $D_u=1$, $D_v=7$. Panel b). Patterns are present when both species share network $A^0$, as previously defined. The patterns are instead impeded  when the network explored by the activator is made sufficiently different from that of the inhibitor, according to formula $A^u(\epsilon)=A^0+\epsilon(A^1-A^0)$.  Here $A^1$ is a realisation of a  Watts-Strogatz~\cite{WS1998} network,  obtained using $N=100$ nodes, $\langle k\rangle=6$ and a probability to rewire of $p=0.6$. The Brusselator model is assumed with parameters $b=8$, $c=22$ and diffusion coefficients $D_u=1$, $D_v=7.1$. In both panels the circles denote the numerically computed, hence exact, dispersion relation. This latter follows from the eigenvalues $\boldsymbol{\mathcal{\tilde{J}}}$, defined in Eq.~\eqref{eq:pert_J}. The solid line (red on line) refers to the first order correction in the perturbative scheme, while the black dot-dashed line is computed from the second order solution given by Eq.~\eqref{eq:pert_lambda}
.}
\label{fig:Fig1}
\end{figure}

The second order correction in Eq.~\eqref{eq:pert_lambda} requires determining the whole spectrum of the perturbed matrix. This information is implicitly stored in
the correction term $\lambda^{(1)}_{max}$. By operating under the additional assumption that $\lambda^{(0)}_{max}$ is much larger than other eigenvalues of the collection, one can obtain a simplified expression for $\tilde{\lambda}_{max}$  which entirely relies on $0$--th order quantities (see derivation in the Appendix~\ref{appendice1})
\begin{equation}
\label{eq:secorde}
\hat{\lambda}_{max}^2= \frac{(\phi^{(0)}_{max})^T \boldsymbol{\mathcal{D}}_0(\lambda^{(0)}_{max}-\boldsymbol{\mathcal{\tilde{J}}}_0)^{-1}\boldsymbol{\mathcal{D}}_0\psi^{(0)}_{max}}{(\phi^{(0)}_{max})^T\cdot \psi^{(0)}_{max}} \, .
\end{equation}
This latter contains as particular case the approximation provided in~\cite{ROH2006,MNN2010}, as discussed in the Appendix.

\begin{figure}[ht!]
\begin{center}
\subfigure[]{
\includegraphics[width=0.55\textwidth]{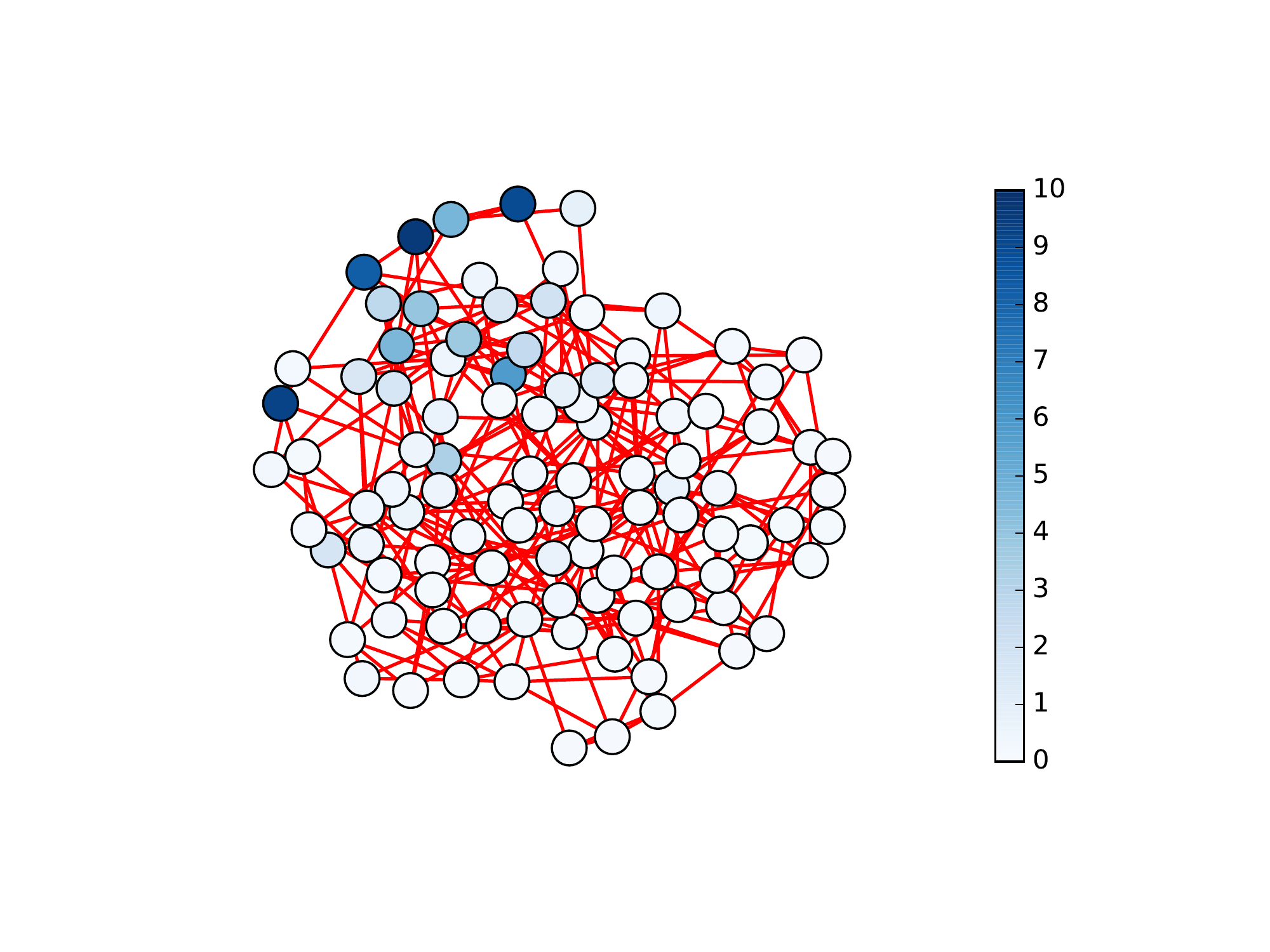}
}\hspace*{-1.5cm}
\subfigure[]{\includegraphics[width=0.55\textwidth]{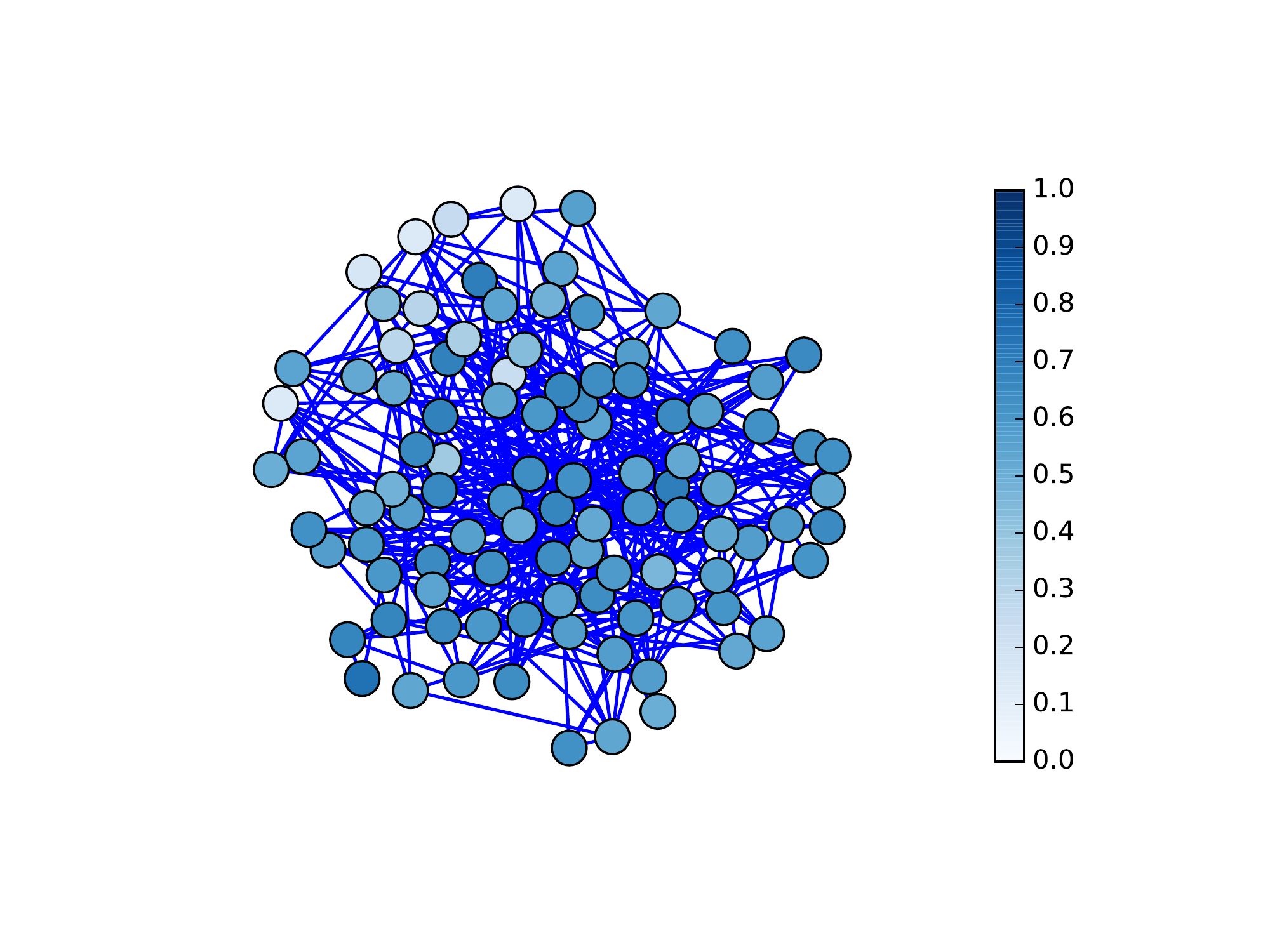}
}
\end{center}
\caption{Patterns in a multigraph. In the panel a) we report the concentration for the activator species and its network (red links) $A^u$. In the panel b) we shown the concentration of the inhibitor with the corresponding network (blue links) $A^v$. In both cases the concentration of the species is normalised to the corresponding equilibrium values (that is $u_i/u^*$ and $v_i/v^*$) and plotted with an apt colour code. The parameters for the Brusselator model are $b=8$, $c=10$ and the diffusion coefficients are $D_u=1$ and $D_v=10$. The networks  $A^u$ and $A^v$ are the same used in Fig.~\ref{fig:Fig1} for the corresponding species with a value of $\epsilon =1$. One can appreciate the differentiation of the nodes content, rich nodes in the amount of $u$ are poor in $v$, and vice-versa.}
\label{fig:Fig2}
\end{figure}

\section{Localised patterns}
\label{sec3}

An interesting property of the reaction-diffusion system on multigraph is the presence of {\em localised patterns}. When the pattern is fully developed, the nodes of the networks can be ideally grouped into two sets: activator (resp. inhibitor) rich (resp. poor) and activator (resp. inhibitor) poor (resp. rich). The relative size of the groups is very different, and one in particular observes that the activators tends to accumulate in a few localised spots.

\begin{figure}[ht!]
\begin{center}
\subfigure[]{
\includegraphics[width=0.5\textwidth]{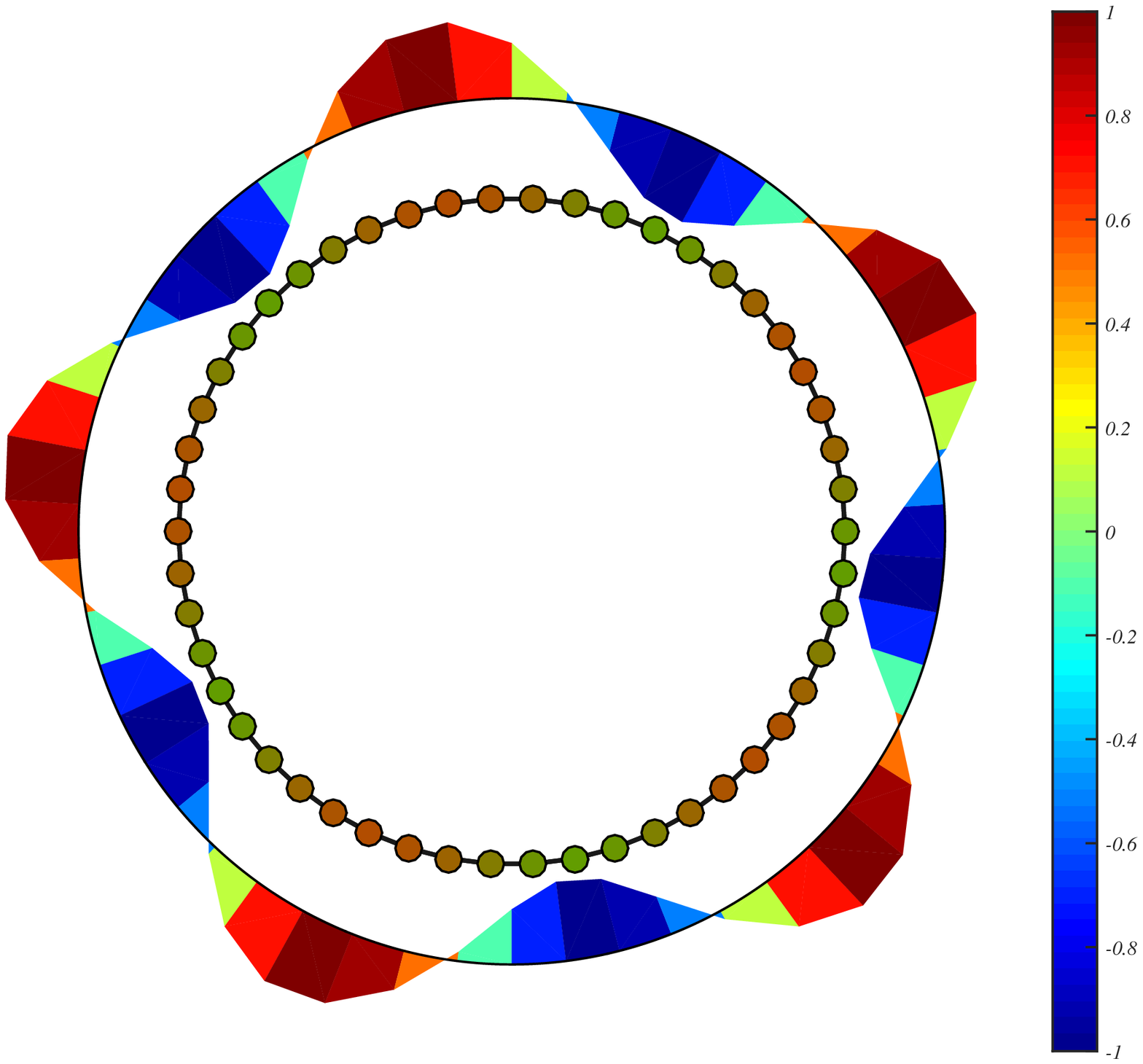}
}\hspace*{-1cm}
\subfigure[]{
\includegraphics[width=0.5\textwidth]{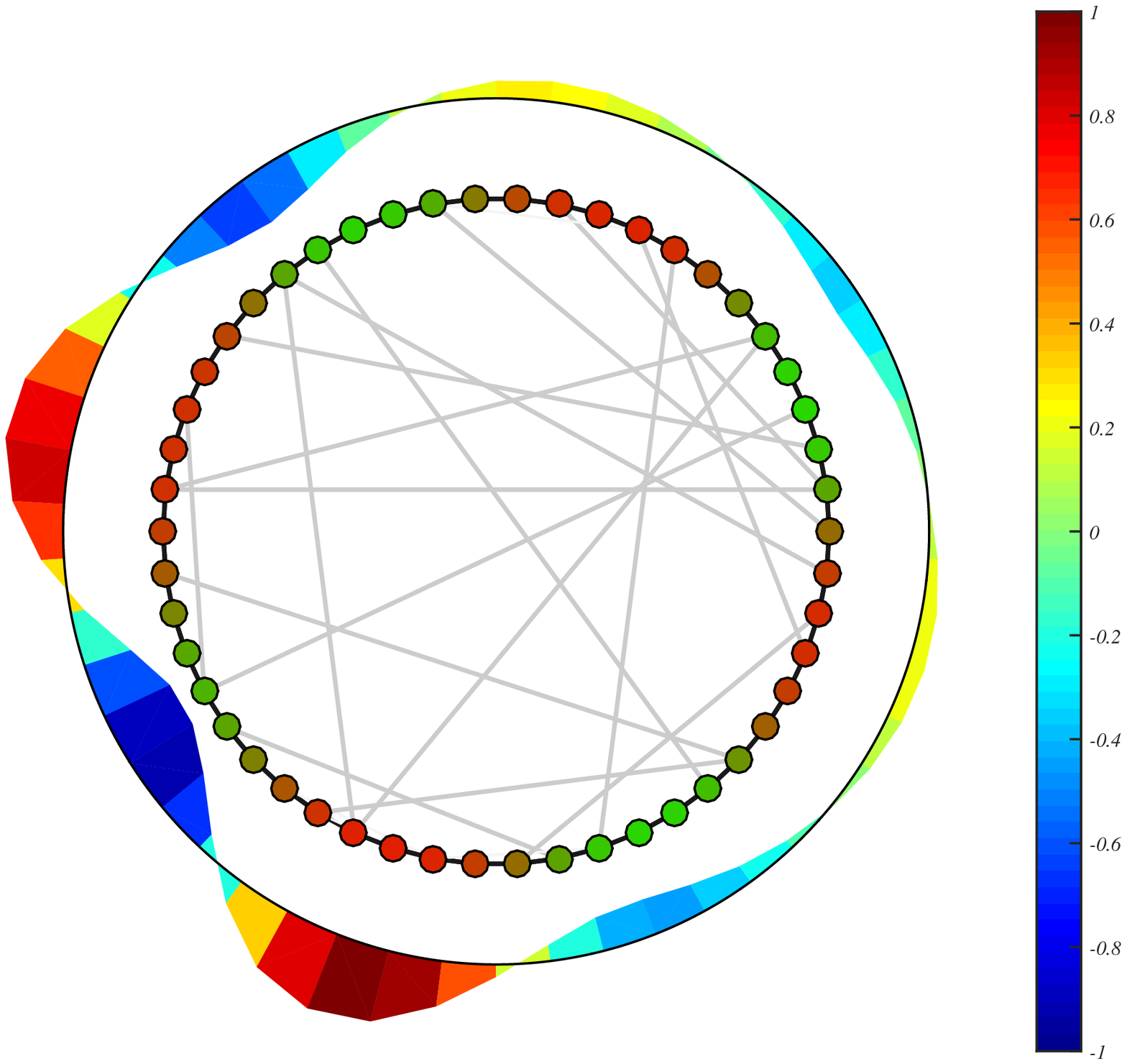}
}\\
\subfigure[]{
\includegraphics[width=0.5\textwidth]{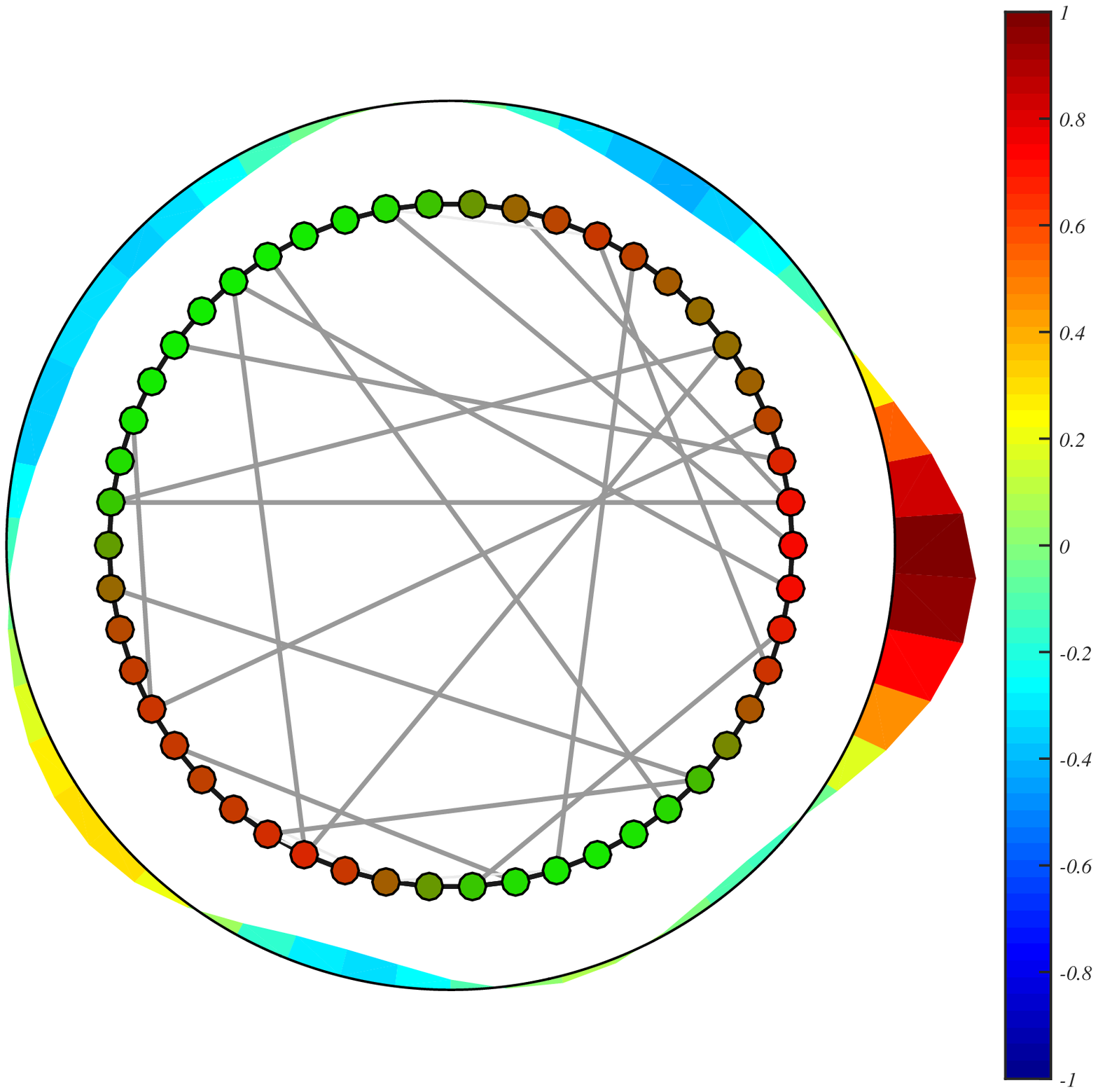}
}\hspace*{-1cm}
\subfigure[]{
\includegraphics[width=0.5\textwidth]{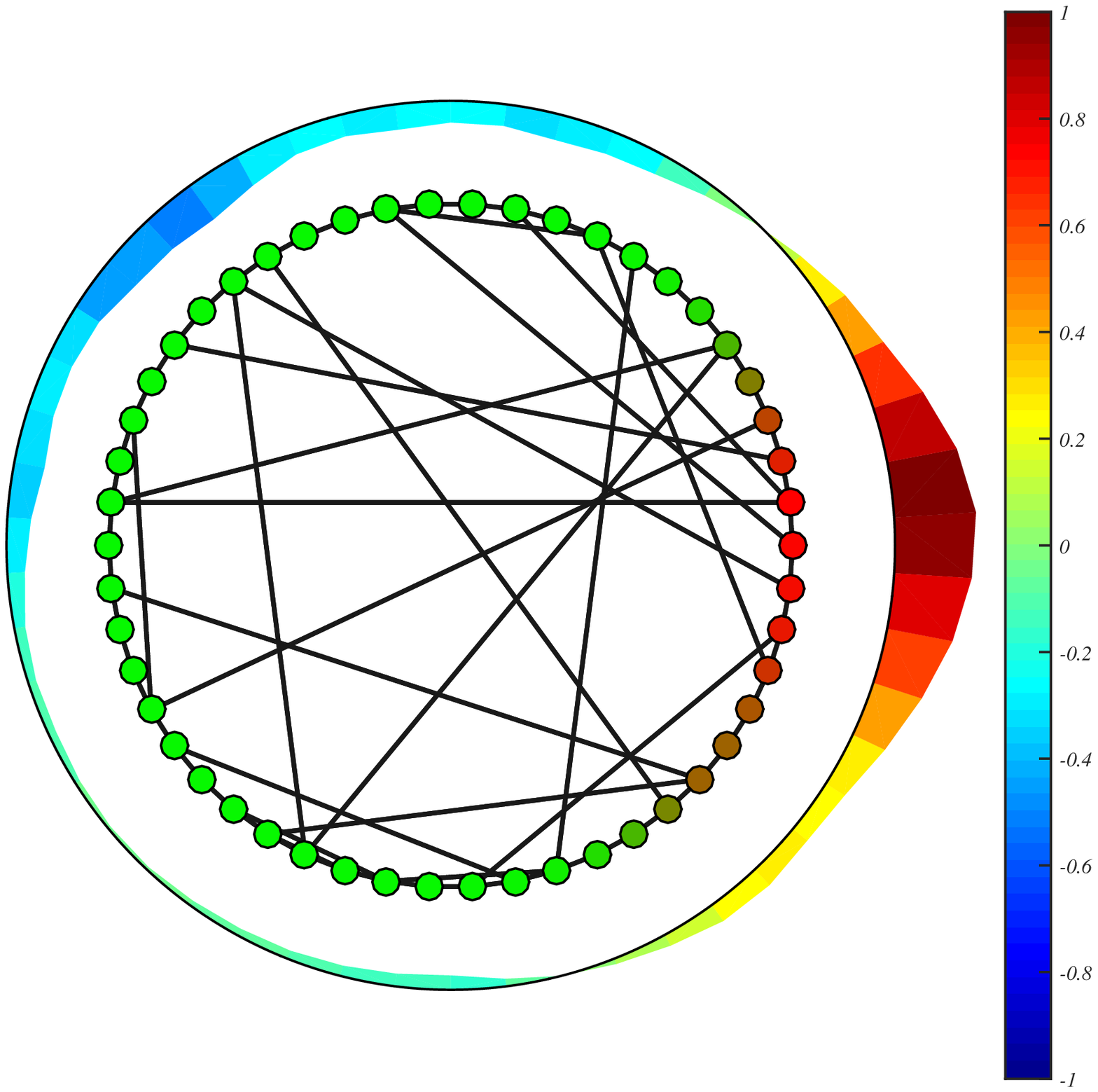}
}
\end{center}
\caption{Localised patterns. The periodic patterns on a $1D$-lattice are destroyed once inhibitors are allowed to use a disordered network to move among the nodes. The activators always diffuse on the regular ring, specified by the adjacency matrix $A^0$. In all panels, we report the asymptotic concentration of activators.
Panel a): The initial configuration, $\epsilon=0$. $A^0$ represents a $1D$-ring. The activator shows a regular pattern, of period $5$ (the same holds true for the inhibitor). Panel b): the inhibitors diffuse on a modified networks $A^v(\epsilon)=A^0+\epsilon (A^2-A^0)$ with $\epsilon=0.05$. Here $A^2$ is a is a Newman-Watts network~\cite{NewmanWatts} with $N=50$ nodes and $p=0.5$, generated starting from a $A^0$. The activator still exhibit periodic patterns but now of period $3$. Panel c): the same as panel b) but with $\epsilon=0.085$. The patterns displayed by activators loose the periodicity and the concentrations start to increase in a few nodes. Panel d):  same as panel b) but with $\epsilon=1$, the patterns are aperiodic and  extremely localised. In all panels we show only the links of the inhibitor network, links with larger weights are shown with darker and thicker lines. The parameters of the Brusselator model are $b=8$, $c=20$, the diffusion coefficients $D_u=8$, $D_v=80$. The outer drawing represents the eigenvector localisation, the black circle denotes the zeroth level, red-yellow colours are associated to positive entries of $\psi_{u,max}$ , while blue-light blue to negative values.}
\label{fig:Fig3}
\end{figure}

To illustrate this phenomenon, we again assume the unperturbed adjacency matrix $A^0$ to identify a ring,  a $1D$-lattice with periodic boundary conditions. We set the parameters so that patterns {\it \`a la Turing} can develop when the two species are diffusing on the same network, as specified by $A^0$,  see Fig.~\ref{fig:Fig3} panel a) where the concentration of 
the activator is depicted on each node.  Patterns reflect the symmetry of the support which defines the spatial backbone of the model: a $5$-fold periodic distribution is observed for this specific case study. When inhibitors are made to evolve on a network which loses progressively its inherent symmetry, as exemplified by the relation 
$A^v(\epsilon)=A^0+\epsilon (A^2-A^0)$,  patterns get steadily localised and thus interest a limited subset of nodes. These facts can be appreciated in Fig.\ref{fig:Fig3} panels b), c) and d), where the  concentrations of the activator species is reported in each nodes, for increasing values of $\epsilon$.

Despite patterns formation is a non--linear process, one can trace back the pattern localisation to the spectral properties of matrix $\mathbf{\tilde{J}}$, which appear to rule the dynamics of the system in the linear regime of the evolution.  Let $\psi_{max}$ be the right eigenvector of $\mathbf{\tilde{J}}$, relative to the most unstable eigenvalue, i.e. the eigenvalue with largest positive real part. Denote with $\psi_{u,max}$ the first $\Omega$ components of $\psi_{max}$, namely the components associated to $u$. The entries of $\psi_{u,max}$ are plotted in Fig.\ref{fig:Fig3} : the external black circle sets the zero. Negative values appear in shades from blue--light to blue and are trapped inside the ring. Positive elements of 
 $\psi_{u,max}$ extend beyond the reference ring and are coloured in red-yellow. The larger $\epsilon$ (i.e. the more disordered the network which hosts the inhibitors) the more localised the eigenvectors~\cite{nakao,McGrawMenzinger,SatorrasCastellano2016}.  A strong correlation is seen between positive components of the most unstable eigenvectors and the region dense in activators. Similarly, negative components of $\psi_{u,max}$ are associated to nodes depleted in activator content. By increasing $\epsilon$ the correlation gets even more significant.
 
\section{Control the instability by inserting a single weighted link}
\label{ssec:singlelink}

In the previous section, we considered the case where all the links are simultaneously, but modestly, changed by varying the value of $\epsilon$. We now turn to consider a different scenario. As before, the two species are initially assumed to evolve on the same network. Then we allow the creation of just one weighted link, in one of the layers. More precisely, $A^u$ and $A^v$ differ due to the insertion of just one individual undirected edge, with weight $w\in[0,1]$. We will hereafter show that such a punctual modification maybe sufficient to induce the creation of patterns, that are impeded when $w$ is set to zero.  Conversely, patterns which can develop when $w=0$ can be deterred when the additional link is switched on. In short, our method  configures as a veritable strategy to control the ability of the system to self-organise at the macroscopic level.  As a matter of fact, we will also determine the nodes which, upon insertion, are predicted to return the most important modification of the pre-existing conditions.

In the following, we shall assume that species $u$ explores an assigned network, characterised by the adjacency matrix
$A^0$. Species $v$ is instead associated to the network described by the adjacency matrix $A^0+T^{(ij)}$, i.e.
the same network on which $u$ is confined  with the additional inclusion of an undirected extra link $(ij)$. Then, Eq.~\eqref{eq:pert_J} rewrites
\begin{equation}
\boldsymbol{\mathcal{\tilde{J}}}=\left( \begin{array}{ccc}
f_u \mathbf{I}_\Omega + D_u\boldsymbol{L}^0 & f_v \mathbf{I}_\Omega\\
g_u \mathbf{I}_\Omega & g_v \mathbf{I}_\Omega + D_v\boldsymbol{L}^0
 \end{array} \right)+\left( \begin{array}{ccc}
\mathbf{0}_\Omega & \mathbf{0}_\Omega\\
\mathbf{0}_\Omega & wD_v\boldsymbol{L}_T
 \end{array} \right)=\boldsymbol{\mathcal{\tilde{J}}}_0+wD_v\boldsymbol{\mathcal{T}}_0\, ,
 \label{eq:pert_Jij}
\end{equation}
where $\boldsymbol{L}_T$ is the Laplacian associated to matrix $T$, whose elements are all identical to zero, except for the element of position $(ij)$, and its symmetric homologue 
$(ji)$ which are set to one. The $2\Omega\times 2\Omega$ matrix $\boldsymbol{\mathcal{T}}_0$ is defined by the last equation. Here, self-loops are not admitted, hence  $i\neq j$. Moreover,  multilinks are not allowed for. This in turn implies that an edge between nodes labeled $i$ and $j$ (the non trivial element of matrix $T$) can be drawn, only if it did not exist in the original formulation of $A^0$, namely $A^0_{ij}=0$.
As anticipated, we are interested in assessing how the newly inserted edge modifies the inherent ability of the system to give rise to  self-organised patterns. The produced effect  will heavily depend on the selected pair of nodes, and their combined topological features. To progress in the analysis we will make use of a perturbative approach to the study of the modified dispersion relation. In particular, the weight $w$ will serve as small parameter in the expansion. The formula reported below is obtained by arresting the expansion, at the first perturbative order. Higher order terms can be accounted for, as follow the strategy outlined in the preceding Section. 

Introduce now the following notation. Given any vector $\vec{x}\in\mathbb{R}^{2\Omega}$, then we define $\vec{x}_u=(x_1,\dots,x_{\Omega})$ and $\vec{x}_v=(x_{\Omega+1},\dots,x_{2\Omega})$ such that $\vec{x}=(\vec{x}_u,\vec{x}_v)$. Hence, the insertion of the undirected link $(ij)$ will materialise in a modified dispersion relation which reads:

\begin{equation}
\tilde{\lambda}_{max}=\lambda^{(0)}_{max}+w\lambda_{max}^1+\mathcal{O}(w^2)=\lambda^{(0)}_{max}+wD_v\frac{(\phi^{(0)}_{v,max})_i(\psi^{(0)}_{v,max})_j}{(\phi^{(0)}_{max})^T\cdot \psi^{(0)}_{max}}+\mathcal{O}(w^2)\, ,
\label{eq:pert_lambdaij}
\end{equation}
that is only the $i$--th component of the $v$--split part of the left eigenvector, $(\phi^{(0)}_{v,max})_i$, and the $j$--th component of the  $v$--splitted part of the right eigenvector, $(\psi^{(0)}_{v,max})_j$, associated to $\lambda^{(0)}_{max}$ enters the first order correction term.

\begin{figure}[ht!]
\begin{center}
\includegraphics[width=0.5\textwidth]{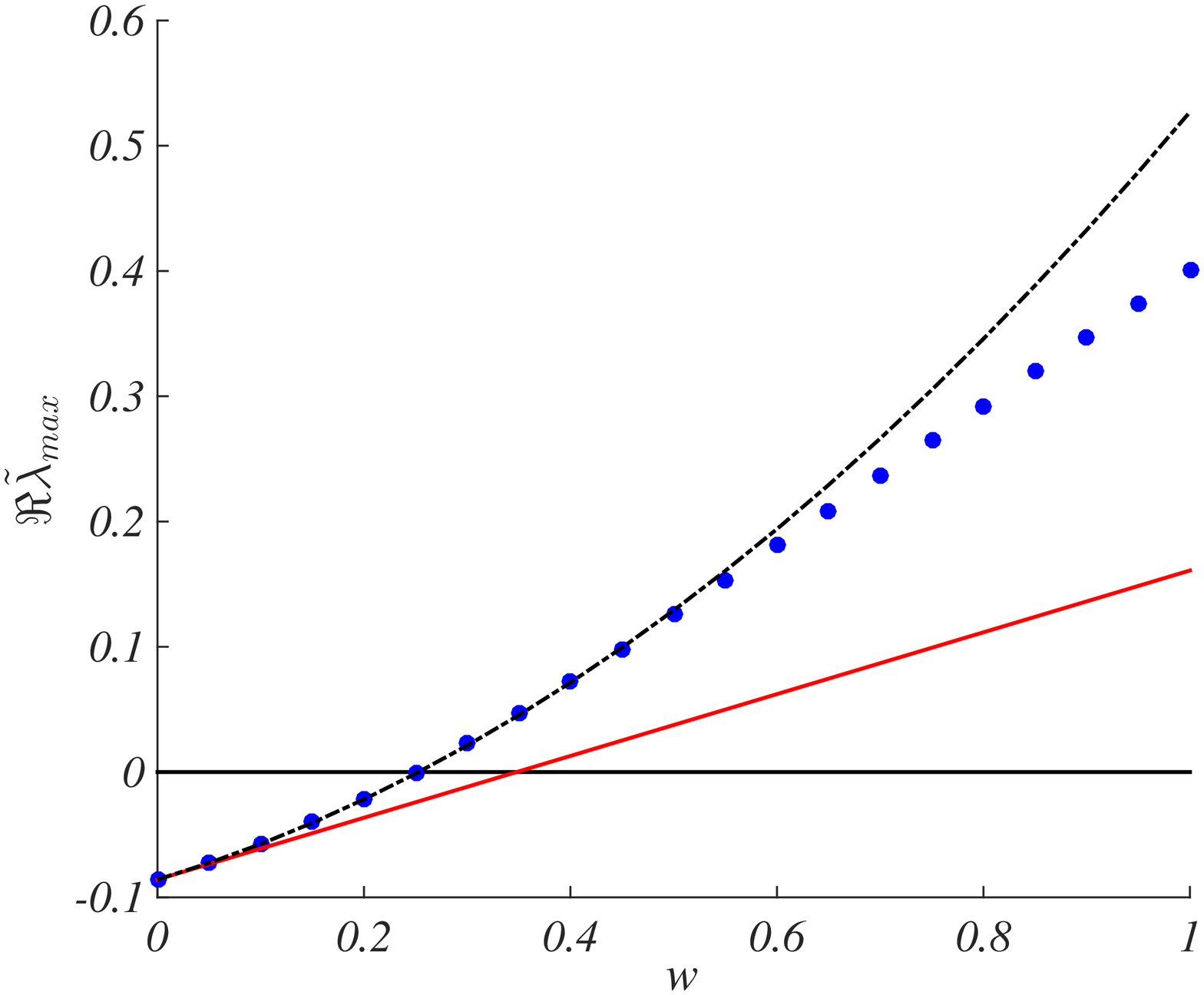}
\end{center}
\caption{Dispersion relation for the multigraph: the original dynamics of the system is controlled via the insertion of a newly added, symmetric and weighted, link, between a pair of nodes $(ij)$. Patterns do not exist when both species employ the same network, with adjacency matrix $A^0$, to explore the available nodes. Here, $A^0$ is generated via the $WS$ recipe: the resulting network is made by $N=50$ nodes; each node is connected to other $6$ nodes of the network ($<k>=6$) with a rewiring probability $p=0.01$. Patterns can instead develop when the 
networks that hosts the inhibitor is altered: the modified adjacency matrix reads $A^v=A^0+wT^{(ij)}$, where only one undirected link $(ij)$ has been added.
The circles denote the exact dispersion relation numerically computed from Eq.~\eqref{eq:pert_J} and plotted as a function of the strength $w$. The solid line (red online) refers to the 
 approximated analytical solution, arrested at the first order of the expansion. The  black dot-dashed line includes also second order corrections.}
\label{fig:reldispFigIJLinks}
\end{figure}

\begin{figure}[ht!]
\begin{center}
\subfigure[]{
\includegraphics[width=0.5\textwidth]{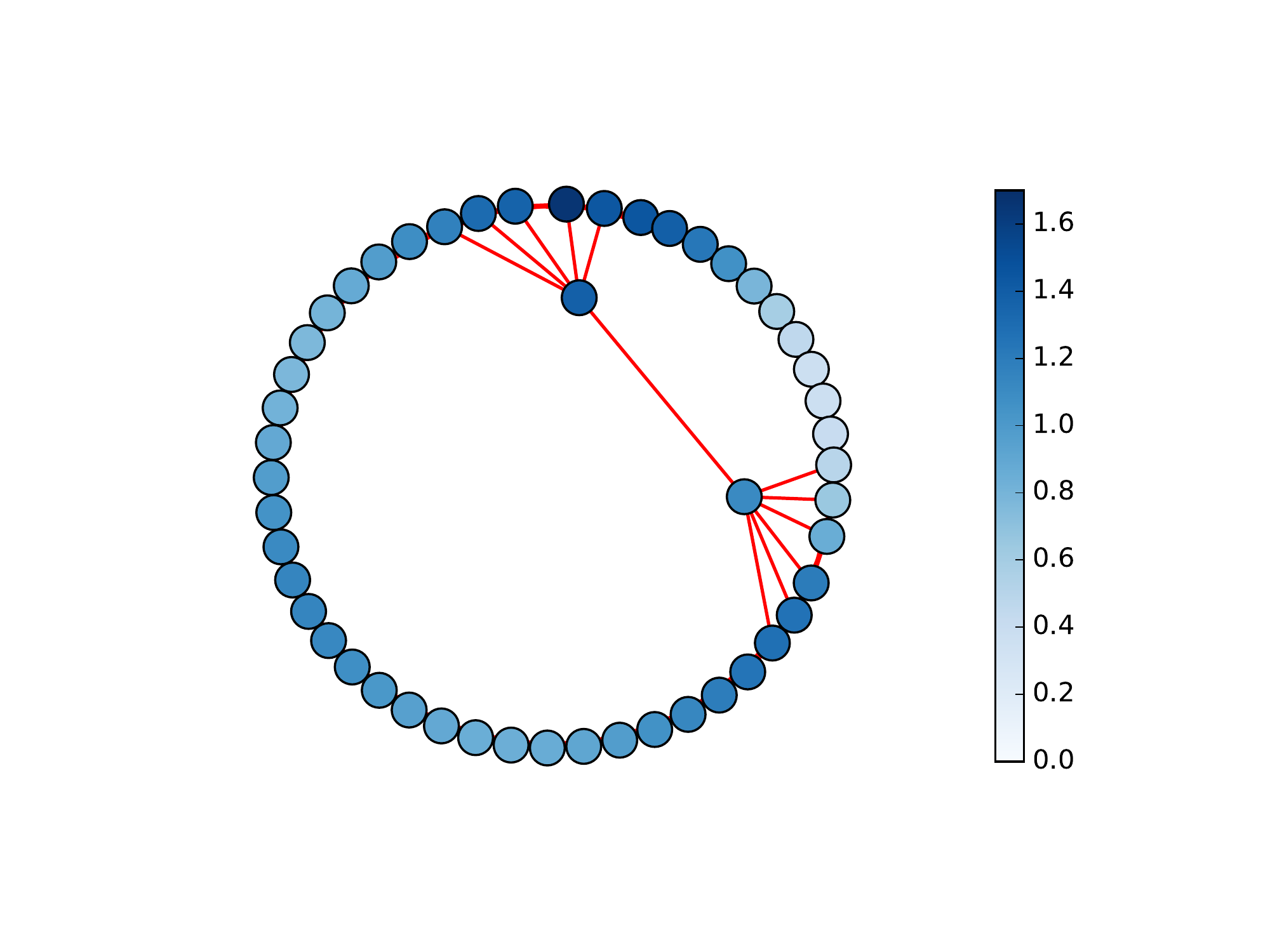}
}\hspace*{-1cm}
\subfigure[]{
\includegraphics[width=0.5\textwidth]{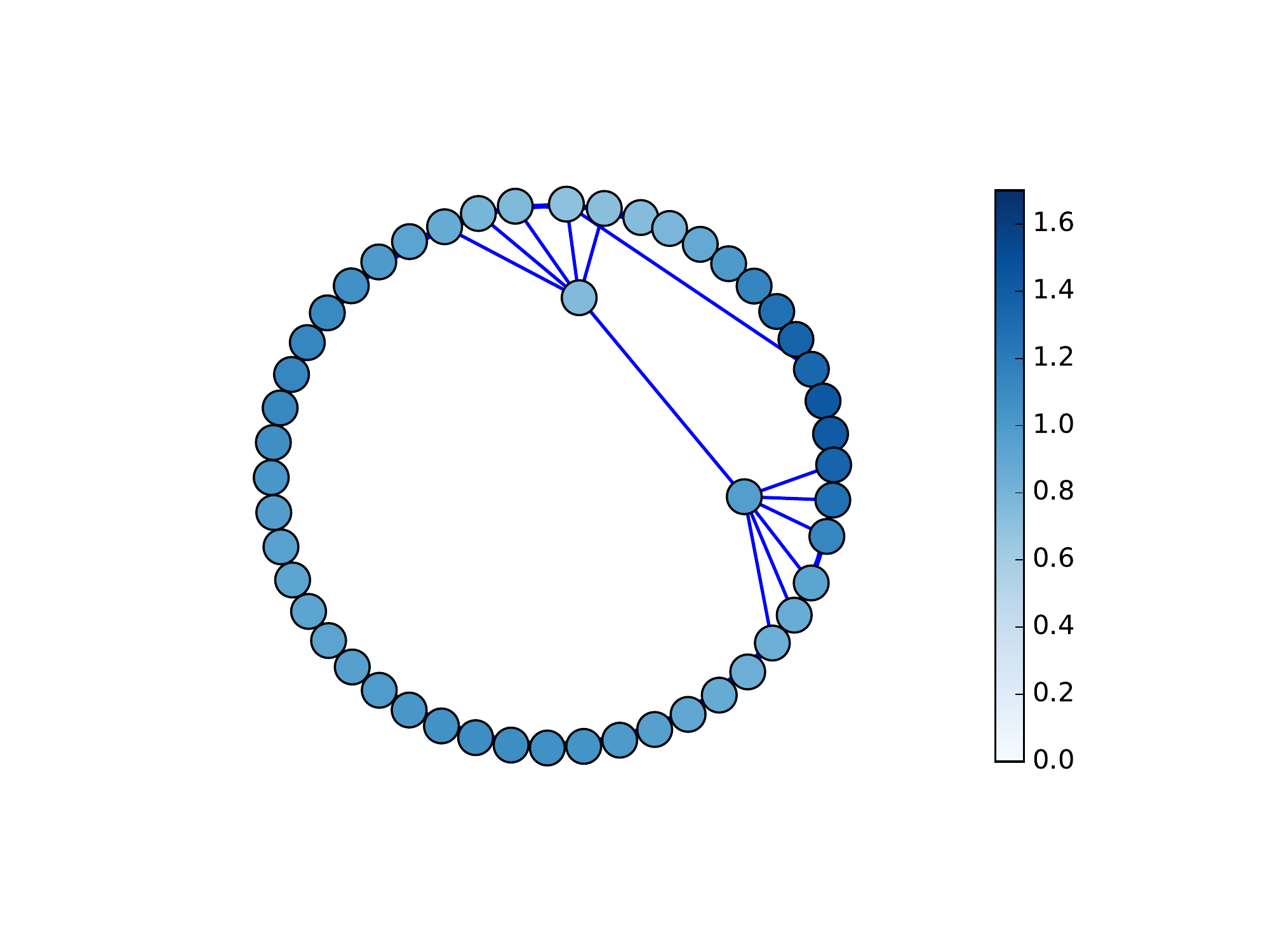}
}
\end{center}
\caption{Creation of patterns by adding one link. $A^0$ is a $WS$ network made by $N=50$ nodes, where each node is connected to other $6$ nodes  ($<k>=6$) and a rewiring probability $p=0.01$. Patterns cannot develop when both species evolve on such a networks, because the dispersion relation returns $\Re \lambda^0_{max}=-0.0858<0$. By adding the link $(ij)$ with weight $w=1$ depicted in the figure and using Eq.~\eqref{eq:pert_lambda} we estimate (at the second order correction), $\Re\tilde{\lambda}_{max}\sim 0.5270>0$, the exact value being $\Re\tilde{\lambda}_{max}^{(true)}=0.4002$. Panel a): the asymptotic distribution of $u$ and its network $A^0$, panel b) the asymptotic distribution of $v$ and its network $A^v=A^0+T^{(ij)}$: the newly added link is clearly visible.}
\label{fig:FigIJLinks}
\end{figure}

Assume that patterns cannot develop for $w=0$, i.e. when both species are evolving on exactly the same network, as specified by the adjacency matrix  $A^0$. 
We have therefore, $\Re \lambda^0_{max}<0$. We also require $\Re \lambda^0_{max}$  to be small enough, for the perturbative scheme devised above to accurately signal possible  instabilities, byproduct of the newly inserted link. By making use of formula Eq.~\eqref{eq:pert_lambda} one can readily determine, which of the selected pair $(ij)$ registers the most relevant increment in the unperturbed, negative, dispersion relation $\Re \lambda^0_{max}$. In other words, we build a symmetric edge between the pair of nodes that are seen to yield the largest $\Re\tilde{\lambda}_{max}>0$. Results of the analysis are reported in Fig.~\ref{fig:reldispFigIJLinks}, where the approximate analytical formula is challenged versus the exact numerical solution for the dispersion relation $\Re\tilde{\lambda}_{max}$ vs. $w$. In Fig.~\ref{fig:FigIJLinks} we display the pattern emerging for the Brusselator model. Here, species $u$ is made to diffuse on a $WS$ network $A^0$ made by $N=50$ nodes with $<k>=6$ and rewiring probability $p=0.01$. The species $v$ diffuses on the network with adjacency matrix $A^v=A^0+T^{(ij)}$, where the additional link $(ij)$ is chosen so to maximise the magnitude of the (positive) first order correction computed in Eq.~\eqref{eq:pert_lambdaij}.

\section{Conclusions}
\label{sec4}
In this paper we studied the problem of patterns formation in a reaction-diffusion model defined on a multigraph network. This representative scenario can be invoked when distinct families of interacting agents employ distinct paths to reach the same target sites, the node of the networks. The aim of the paper is to analyse how small changes in the topological characteristics of the networks may interfere with the intrinsic ability of the system to yield self-organised patterns, via a symmetry breaking instability of the Turing type.
Assuming that the modifications imposed can be traced back to a small scalar parameter, we obtained approximated analytical formulae for the relevant dispersion relation. We can hence foresee the effect of any enforced topological change to the structure of the hosting support, so yielding a machinery which could possibly guide the definition of novel  protocols to control the dynamics of the systems on networks \cite{control1, control2}.  We have in fact shown that patterns can be created (or, alternatively, destroyed) by tuning the characteristic of the networks 
on which the dynamics take place.  Importantly, the macroscopic dynamics can be shaped as sought by adding/deleting one individual weighed link, which therefore acts as a veritable dynamical switcher. We also studied the phenomenon of patterns localisation and explained it in terms of the eigenvector localisation.

\section*{Acknowledgments}
The work of T.C. and M.A. presents research results of the Belgian Network DYSCO (Dynamical Systems, Control, and Optimization), funded by the Interuniversity Attraction Poles Programme, initiated by the Belgian State, Science Policy Office. D.F. acknowledges financial support of the program Prin 2012 financed by the Italian Miur.

\appendix
\section{Eigenspectrum perturbation method}
\label{appendice1}

The eigenvalues problem~\eqref{eq:pert_J} can be cast in the following general form. Given a matrix $A_0$, whose eigenvalues and eigenvectors are known, and a small parameter $\epsilon$, we wish to determine the eigenvalues of the matrix $A_0+\epsilon A_1$. $A_1$ acts as a perturbation rescaled by the small scalar quantity $\epsilon$.

Let us introduce the matrices $\Lambda(\epsilon)=diag(\lambda_{1}(\epsilon),\lambda_{2}(\epsilon), \ldots \lambda_{2\Omega}(\epsilon))$  and $\Psi(\epsilon)=\left(\begin{matrix} \vec{\psi}_1(\epsilon)&
\vec{\psi}_2(\epsilon) &
\ldots &  \vec{\psi}_{2\Omega}(\epsilon) \end{matrix}\right)$ and 
expand them in powers of $\epsilon$ as follows:
\begin{equation}
\label{ps}
\Lambda(\epsilon)=\sum_{l\geq 0} \Lambda_l \epsilon^l \quad\text{and}\quad \Psi(\epsilon)=\sum_{l\geq 0} \Psi_l \epsilon^l\, ,
\end{equation}
where $\Lambda_0$ stands for the eigenvalues of the unperturbed matrix $A_0$; $\Psi_0$ (resp. $\Phi_0$, to be used later) stands for the matrix whose columns (resp. rows) are the right (resp. left) eigenvectors of  $A_0$, $\vec{\phi}_{0,i}$ (resp. $\vec{\psi}_{0,i}$). Inserting formulae~\eqref{ps} into the eigenvalues problem for the perturbed system $(A_0+\epsilon A_1)\Psi=\Psi\Lambda$ and collecting together the terms of same order in $\epsilon$ beyond the trivial zero{-th} order contribution, we get
\begin{equation}
\label{eq:pertsys2}
A_0\Psi_l+A_1\Psi_{l-1}=\sum_{k=0}^l\Psi_{l-k}\Lambda_k{\quad\forall l\geq 1}\, .
\end{equation}
{Left} multiplying the previous equation by $\Phi_0$ and setting $C_l=\Phi_0 \Psi_l$ yields:
\begin{equation}
\label{eqCLambda}
\Lambda_0C_l-C_l\Lambda_0=-\Phi_0A_1\Psi_{l-1}+C_0\Lambda_l+\sum_{k=1}^{l-1}C_{l-k}\Lambda_k\, .
\end{equation}

Eq.~\eqref{eqCLambda} contains two unknowns, namely $C_l$ and $\Lambda_l$ for all $l\geq 1$. To obtain the analytical solution used in the main text we observe that Eq.~\eqref{eqCLambda} can be cast in the compact form
\begin{equation}
\label{eq:operator}
[\Lambda_0,X]=Y\, ,
\end{equation}
where $X$ and $Y$ are $2\Omega\times2\Omega$ matrices and $[\cdot,\cdot]$ stands for the matrix commutator. In practice, given $Y\in \mathbb{R}^{2\Omega\times2\Omega}$,  one needs to find $X\in \mathbb{R}^{2\Omega\times2\Omega}$ solution of~\eqref{eq:operator}. Since 
$\Lambda_0$ is a diagonal matrix, the codomain of the operator $[\Lambda_0,\cdot]$ is formed by all the matrices with zero diagonal. To self-consistently solve~\eqref{eq:operator} it is therefore necessary to impose that $Y$ has zero diagonal elements. Therefore the matrix $X$ will have its diagonal elements undetermined. 

Because of the above remark one can solve Eq.~\eqref{eqCLambda} by setting $\Lambda_l$ so to cancel the diagonal terms on its right hand side, that is:
\begin{equation}
\label{eq:fixlambda1}
(\Lambda_l)_{ii} =\frac{(\Phi_0 A_1 \Psi_{l-1})_{ii}-\sum_{k=1}^{l-1}(C_{l-k}\Lambda_k)_{ii}}{\vec{\phi}^T_{0,i}\cdot \vec{\psi}_{0,i}} \quad \text{and }(\Lambda_l)_{ij}=0\text{ for $i\neq j$}\,,
\end{equation}
where the denominator is nothing but $(C_0)_{ii}$. Assuming the eigenvalues of $A_0$ to be simple, then $C_l$ is readily found to match:
\begin{equation}
\label{eq:fixB}
(C_l)_{ij} = \frac{(-\Phi_0 A_1 \Psi_{l-1})_{ij}+\sum_{k=1}^{l-1}(C_{l-k}\Lambda_k)_{ij}}{\lambda^{(0)}_i-\lambda^{(0)}_j} \text{ if $i\neq j$} \quad (C_l)_{ii}=0\, .
\end{equation}
This latter epression allows us to simplify~\eqref{eq:fixlambda1}.  In fact:
\begin{equation*}
(C_{l-k}\Lambda_k)_{ii}=\sum_h(C_{l-k})_{ih}(\Lambda_k)_{hi}=0\, ,
\end{equation*}
and thus the approximated eigenvalues are given by
\begin{equation}
\label{eq:fixlambda}
(\Lambda_l)_{ii} =\frac{(\Phi_0 A_1 \Psi_{l-1})_{ii}}{\vec{\phi}^T_{0,i}\cdot \vec{\psi}_{0,i}} \quad \text{and }(\Lambda_l)_{ij}=0\text{ for $i\neq j$}\,.
\end{equation}
In the case $l=1$ the previous formula reduces to:
\begin{equation}
\label{eq:casel2}
\lambda^{(1)}_i =\frac{\vec{\phi}^T_{0,i} A_1 \vec{\psi}_{0,i}}{\vec{\phi}^T_{0,i}\cdot \vec{\psi}_{0,i}} \quad \text{and}\quad (C_1)_{ij} = -\frac{\vec{\phi}^T_{0,i} A_1 \vec{\psi}_{0,j}}{\lambda^{(0)}_i-\lambda^{(0)}_j} \quad \text{for $i\neq j$,}
\end{equation}
which is the first order approximation used in the main text. 

The second order correction can also be straightforwardly obtained:
\begin{equation}
\label{eq:secord}
\lambda^{(2)}_i=\frac{(\Phi_0 A_1\Psi_{1})_{ii}}{\vec{\phi}^T_{0,i}\cdot \vec{\psi}_{0,i}}=\frac{(\Phi_0 A_1\Phi_0^{-1}C_{1})_{ii}}{\vec{\phi}^T_{0,i}\cdot \vec{\psi}_{0,i}}\, ,
\end{equation}
where we used the definition $C_1=\Phi_0\Psi_1$. Observe that $C_1$ will involve all the eigenvectors and so it does the second order correction. Using the explicit form of $C_1$ given by Eq.~\eqref{eq:casel2} Eq.~\eqref{eq:secord} can be rewritten as:
\begin{equation}
\label{eq:secordb}
\lambda^{(2)}_i=-\frac{1}{\vec{\phi}^T_{0,i}\cdot \vec{\psi}_{0,i}}\sum\frac{(\Phi_0)_{im}(A_1)_{mn}(\Phi_0^{-1})_{nk}(\Phi_0)_{kp}(A_1)_{pq}(\Psi_0)_{qi}}{\lambda^{(0)}_k-\lambda^{(0)}_i}\, ,
\end{equation}
where the sum runs over all the repeated indexes.

Assuming now $\lambda^{(0)}_i$ to be the largest eigenvalue, namely $\lambda^{(0)}_i>>\lambda^{(0)}_k$ for all $k\neq i$,  using the expansion $1/(1-x)=\sum_{l\geq 0}x^l$ for $x=\lambda^{(0)}_k/\lambda^{(0)}_i$ and recalling that $(\lambda^{(0)}_k)^l (\Phi_0)_{kp}=(\Lambda_0^l\Phi_0)_{kp}=(\Phi_0A_0^l)_{kp}$, we can rewrite the previous formula as follows:
\begin{eqnarray}
\label{eq:secordd}
\lambda^{(2)}_i&=&\frac{1}{\vec{\phi}^T_{0,i}\cdot \vec{\psi}_{0,i}}\sum_{l\geq 0}\sum \frac{1}{(\lambda^{(0)}_i)^{l+1}}{(\Phi_0)_{im}(A_1)_{mn}(\Phi_0^{-1})_{nk}(\Phi_0A_0^l)_{kp}(A_1)_{pq}(\Psi_0)_{qi}}=\notag\\
&=&\frac{1}{\vec{\phi}^T_{0,i}\cdot \vec{\psi}_{0,i}}\sum_{l\geq 0} \frac{1}{(\lambda^{(0)}_i)^{l+1}}(\Phi_0A_1A_0^lA_1\Psi_0)_{ii}=\frac{1}{\vec{\phi}^T_{0,i}\cdot \vec{\psi}_{0,i}}(\Phi_0A_1(\lambda^{(0)}_i-A_0)^{-1}A_1\Psi_0)_{ii}\notag\\
&=&\frac{\vec{\phi}^T_{0,i}A_1(\lambda^{(0)}_i-A_0)^{-1}A_1\vec{\psi}_{0,i}}{\vec{\phi}^T_{0,i}\cdot \vec{\psi}_{0,i}}\, .
\end{eqnarray}
Observe that the first term, i.e. neglecting all the terms in the sum but $l=0$, is given by
\begin{equation}
\label{eq:secordc}
\lambda^{(2)}_i \sim\frac{1}{\lambda^{(0)}_i} \frac{\vec{\phi}^T_{0,i}(A_1)^2\vec{\psi}_{0,i}}{\vec{\phi}^T_{0,i}\cdot \vec{\psi}_{0,i}}\, ,
\end{equation}
which is nothing but the formula proposed in~\cite{ROH2006,MNN2010}.

\end{document}